\documentclass[11pt,a4paper]{article}

\usepackage[a4paper,margin=2.5cm]{geometry}
\usepackage[utf8]{inputenc}

\usepackage{subfigure,graphicx}
\usepackage{amsmath,amsfonts,latexsym,amssymb,euscript,xr, mathtools}
\usepackage{booktabs}
\usepackage[table]{xcolor}
\usepackage{color,colortbl,tabularx}

\usepackage[english]{babel}
\usepackage[protrusion=true,expansion=true]{microtype}
\usepackage{amsthm}
\usepackage{pifont}
\usepackage[colorlinks=true,linkcolor=blue,citecolor=blue,urlcolor=blue]{hyperref}

\definecolor{LightBlue}{rgb}{0.88,0.9,0.9}

\newcommand\blfootnote[1]{%
  \begingroup
  \renewcommand\thefootnote{}\footnote{#1}%
  \addtocounter{footnote}{-1}%
  \endgroup
}

\title{\bf Radiomics-Based Thyroid Nodule Classification with Label-Conditional Conformal Prediction}

\author{Laura Testa$^{*,1}$, Pierpaolo Brutti$^{1}$, and Umberto Ferraro Petrillo$^{1}$\\[8pt]
{\small $^1$ Department of Statistical Sciences, Sapienza University of Rome, Italy.}\\[6pt]
{\small ORCID codes: LT 0009-0002-8992-157X; PB 0000-0003-0458-7868; UFP 0000-0002-4308-5126.}\\[6pt]
{\small $^*$corresponding author: laura.testa@uniroma1.it}}

\date{}
\begin{document}

\maketitle
\blfootnote{Accepted at the CIBB 2026 conference (\url{https://cibb2026.teralab.ai/}).}

\begin{abstract}
Thyroid nodule malignancy assessment relies on ultrasound imaging, 
yet current risk stratification systems suffer from substantial 
interobserver variability, contributing to a high rate of unnecessary biopsies.
We propose a radiomics-based framework combining quantitative 
feature extraction via PyRadiomics with a 
Generalized Additive Model (GAM) classifier, selected among six candidates for its 
favorable trade-off between discriminative performance and 
interpretability. To complement point predictions with uncertainty estimates, Cross-Conformal Prediction with label-conditional guarantees is applied, flagging ambiguous cases for clinical review rather than forcing a binary decision.
Experiments are conducted on TN5000, a publicly available ultrasound dataset of thyroid nodules.
After Spearman correlation filtering, 36 features 
are retained and the GAM achieves an AUC of $0.89$, a sensitivity of $0.83$, and a specificity of $0.82$ on the validation set. At a significance level 
$\epsilon=0.09$, selected to ensure empirical coverage guarantees 
on both classes, 63\% of nodules on the test set receive a certain (i.e., single-label)
prediction, with conditional precision of $0.94$ for certain malignant 
and conditional Negative Predictive Value (NPV) of $0.73$ for certain benign cases, while the 
remaining 37\% are referred to the clinician. 

\medskip
\noindent\textbf{Keywords:} Thyroid Nodule Classification, Ultrasound Imaging, Radiomics, Generalized Additive Model, Conformal Prediction
\end{abstract}

\section{Introduction}
\label{sec:SCIENTIFIC-BACKGROUND}

Thyroid cancer is one of the most prevalent endocrine malignancies, with an increasing incidence worldwide~\cite{biopsies_rate}. 
Diagnosis relies primarily on ultrasound (US) imaging, which is used to assess nodule characteristics and guide biopsy decisions. However, current practice leads to a high rate of unnecessary biopsies, with at least half of those performed proving to be benign~\cite{biopsies_rate}. Existing risk stratification systems such as ACR TI-RADS~\cite{ti_rads} rely on qualitative visual assessment, introducing interobserver variability and limiting diagnostic reliability.

Radiomics~\cite{radiomics} has emerged as a promising complementary 
approach, enabling operator-independent extraction of quantitative 
imaging features from US images. Combined with machine learning classifiers, radiomics-based models have shown competitive diagnostic performance for thyroid nodule classification~\cite{cleere2022}, with tools such as PyRadiomics~\cite{pyradiomics} providing standardized and reproducible feature extraction in this domain~\cite{salmanpour}.
However, two critical limitations persist in existing approaches: limited interpretability, and the absence of predictive uncertainty quantification, both essential for clinical adoption.

To address these limitations, we propose a framework that 
evaluates multiple machine learning classifiers on radiomic features 
extracted via PyRadiomics from the publicly available 
TN5000 dataset~\cite{tn5000}, selecting the model that best balances 
discriminative performance and interpretability. We further integrate 
Cross-Conformal Prediction with label-conditional coverage 
guarantees to provide formal uncertainty quantification~\cite{vovk2015}, 
allowing the model to explicitly flag diagnostically uncertain cases 
rather than forcing a hard binary classification. 

\section{Data and Methods}
\label{sec:DATA-AND-METHODS}

\subsection{TN5000 Dataset}
\begin{figure}[h]
\centering
\subfigure[Benign]{
    \includegraphics[width=0.48\textwidth]{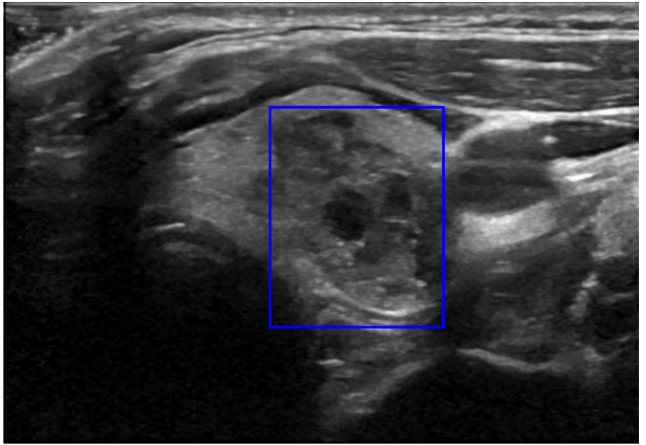}
}
\hfill
\subfigure[Malignant]{
    \includegraphics[width=0.48\textwidth]{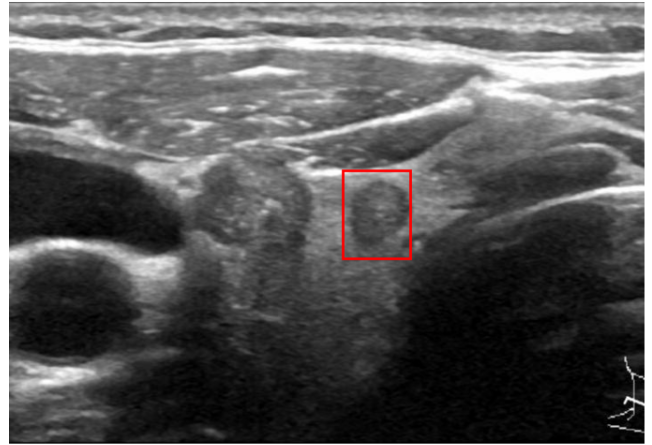}
}
\caption{\textbf{Representative US images from TN5000.} 
Bounding boxes indicate the annotated nodule ROI 
used for radiomic feature extraction (blue: benign, red: malignant).}
\label{fig:sample_pair}
\end{figure}

All experiments were conducted on TN5000~\cite{tn5000}, a publicly 
available dataset comprising 5000 US images of thyroid 
nodules, annotated with biopsy-confirmed labels and bounding box coordinates by expert radiologists.
Focusing exclusively on classification, images were cropped to the bounding box annotations prior to feature extraction, restricting radiomic feature computation to the nodule 
region of interest (ROI), as  shown in~\autoref{fig:sample_pair}.

We adopted the official training/validation/test split provided with the 
dataset, yielding 3500 training, 500 validation, and 1000 test images, 
with approximately 71\% malignant and 29\% benign cases in the 
training set, which was taken into account during model training.

\subsection{Radiomic Feature Extraction}
Radiomic features were extracted from each nodule ROI using 
PyRadiomics v3.1.0~\cite{pyradiomics}, an open-source Python 
library for standardized radiomic feature computation.
Since all images are two-dimensional US scans, extraction was configured in 2D mode with prior intensity normalization and a bin width of 25 for discretization, consistent with the PyRadiomics default.
A total of 102 features were extracted 
from seven feature families: First Order Statistics (18), 
Shape-based 2D (9), Gray Level Co-occurrence Matrix (GLCM, 24), 
Gray Level Run Length Matrix (GLRLM, 16), Gray Level Size Zone 
Matrix (GLSZM, 16), Neighbouring Gray Tone Difference Matrix 
(NGTDM, 5), and Gray Level Dependence Matrix (GLDM, 14).

\subsection{Feature Selection and Standardization}
Given the well-known redundancy of radiomic features~\cite{correlation}, we applied a Spearman correlation-based filter, which captures monotonic relationships without assuming linearity.
For each pair of features with $|\rho| > 0.90$, one was removed~\cite{spearman}, yielding 36 features from five 
of the original seven families: First Order Statistics (12), GLCM (12), 
GLRLM (7), NGTDM (3), and Shape-based 2D (2). Features from the 
GLSZM and GLDM families were entirely removed, indicating high redundancy with the retained ones.
The retained features were then standardized using z-score normalization applied separately for each feature, with mean and standard deviation estimated on the training set and applied to validation and test sets.

\subsection{Classification}
Six classifiers were evaluated on the retained 36 features: Logistic Regression with LASSO penalization (LR-LASSO), 
Generalized Additive Model (GAM), Random Forest, Extra Trees, 
XGBoost, and a Multi-Layer Perceptron (MLP) with two hidden layers. All models were trained with class weights proportional to the 
inverse class frequency to address class imbalance, and subsequently 
compared on the validation set using the Area Under the ROC Curve (AUC), sensitivity, specificity, 
precision, negative predictive value (NPV), and F1-score.


\subsection{Label-Conditional Cross-Conformal Prediction}
To quantify predictive uncertainty, we applied Cross-Conformal 
Prediction (CCP)~\cite{vovk2015}, a framework that produces 
prediction sets with guaranteed coverage $1-\epsilon$ without 
distributional assumptions. 
CCP extends split conformal prediction via $K$-fold cross validation:
the classifier is trained $K$ times on $K-1$ folds, conformity scores are 
computed on the held-out fold, using the full training set as calibration set,  which in our case corresponds to the merged training and validation sets (4000 US images).
Following~\cite{vovk2015}, the computation of the conformity scores is based on the log-odds of the classifier output:
\begin{equation}
    \alpha = \left\{
\begin{array}{crl} 
\log\frac{\hat{P}(Y=1\mid X)}{\hat{P}(Y=0\mid X)} & if & y = 1 \\[6pt]
-\log\frac{\hat{P}(Y=1\mid X)}{\hat{P}(Y=0\mid X)} & if & y = 0
\end{array}
\right.
\end{equation}
Given the class imbalance in TN5000, we adopted the \textit{Label-Conditional} 
Cross-Conformal Prediction (CCCP),
which guarantees coverage separately for each class:
$P(y \in \Gamma^\epsilon \mid Y=y) \geq 1-\epsilon$. The prediction set $\Gamma^\epsilon$ for each 
test image is defined as:
\begin{equation}
\Gamma^{\epsilon} \coloneqq \{y \in \{0,1\} \mid p^y > \epsilon\},
\end{equation}
where the \textit{label-conditional} p-values $p^y$ are:
\begin{equation}
p^y = \frac{\sum_{k=1}^{K} \left|\left\{ i \in S_k \mid y_i = y,\; 
\alpha_{i,k} \leq \alpha^y_k \right\}\right| + 1}
{\sum_{k=1}^{K} \left|\left\{ i \in S_k \mid y_i = y 
\right\}\right| + 1},
\end{equation}
where $\alpha_{i,k}$ is the conformity score of calibration example $i$ 
in fold $S_k$, computed using the model trained on all remaining 
folds $S_{-k}$; $\alpha^y_k$ is the conformity score of the test 
example under assumed label $y$, computed using the same model. In our binary setting, this yields four possible prediction sets for each test image: certain 
benign $\{B\}$, certain malignant $\{M\}$, uncertain $\{B,M\}$, 
or empty $\{\}$. 

\section{Results}
\label{sec:RESULTS}

\subsection{Classifier Selection}
The performance metrics of the six classifiers on the validation set are reported in~\autoref{tab:RES_VALIDATION}.
All models achieved comparable AUC values ($\approx0.88-0.89$), consistent with the performance reported in the radiomics literature 
for thyroid nodule classification~\cite{cleere2022}. 
However, substantial differences emerged in sensitivity and specificity. Tree-based ensemble methods and 
MLP achieved high sensitivity (up to 0.95 for Extra Trees) and NPV at the 
cost of low specificity (as low as 0.51), indicating a tendency 
to over-predict the predominant class of malignancy despite class weighting. XGBoost partially mitigated this imbalance 
($0.72$ specificity), but remained inferior to LR-LASSO and GAM, which achieved the best sensitivity-specificity trade-off. 
\begin{table}[httb!] \small
\centering
\caption{\textbf{Classifier comparison on the validation set.} Performance metrics on the six evaluated models; bold values indicate the best performance per metric.
\label{tab:RES_VALIDATION}}
    \begin{tabularx}{0.82\linewidth}{ l  c  c c c c c }
        \toprule
          \textbf{Model} & \textbf{AUC} & \textbf{Sensitivity}  & \textbf{Specificity} & \textbf{Precision} & \textbf{NPV} & \textbf{F1}\\
        \midrule
        \rowcolor{LightBlue} LR-LASSO  &       0.88 &  0.80 &  0.78  & 0.91  & 0.56  & 0.85\\
         GAM  &     \textbf{0.89}  & 0.83  & \textbf{0.82} &  \textbf{0.93}  & 0.62 &  0.88 \\
        \rowcolor{LightBlue}  XGBoost &         0.88  & 0.85 &  0.72 &  0.90  & 0.62 &  0.88 \\
        Random Forest &    0.88 &  0.94 &  0.55  & 0.86  & 0.76 &  0.90 \\
        \rowcolor{LightBlue} Extra Trees  &     0.88 &  \textbf{0.95} &  0.51 &  0.85 &  \textbf{0.77} &  0.90 \\
        MLP  &      \textbf{0.89}  &  0.93 &   0.62  &  0.88  &  0.75 &   \textbf{0.91}\\
        \bottomrule
    \end{tabularx}
\end{table}
Between the two, the GAM outperformed LR-LASSO in all metrics: sensitivity (0.83 vs 0.80), specificity (0.82 vs 
0.78), precision (0.93 vs 0.91), and NPV (0.62 vs 0.56). 
The GAM was therefore selected as the final classifier for its 
best sensitivity-specificity trade-off and native interpretability, 
modeling the log-odds of malignancy as a sum of smooth spline 
functions of individual features~\cite{GAM}.

\subsection{GAM Feature Importance and Interpretation}
Unlike black-box models, the GAM allows direct inspection of 
each feature's contribution to the predicted malignancy probability. To identify the most influential features, we applied permutation 
importance, a model-agnostic method that measures the decrease in 
predictive performance when each feature is randomly permuted. 
AUC was used as the performance metric over 50 repetitions, given the class imbalance in the dataset, 
with results reported in~\autoref{tab:PERMUTATION}.
\begin{table}[httb!] \small
\centering
  \caption{\textbf{Permutation importance of the top 10 features.} 
 Features are ranked in descending order of mean AUC decrease 
over 50 repetitions; standard deviation is reported across repetitions.
    \label{tab:PERMUTATION}}
    \begin{tabularx}{\linewidth}{ l c c }
        \toprule
          \textbf{Feature} & \textbf{Mean AUC Decrease} & \textbf{Standard Deviation} \\
        \midrule
        \rowcolor{LightBlue} Mean (First Order) & 0.297 & 0.018 \\
       Major Axis Length (Shape2D) & 0.258 & 0.022 \\
        \rowcolor{LightBlue} Entropy (First Order) & 0.251 & 0.018 \\
        Autocorrelation (GLCM)& 0.237 & 0.017 \\
        \rowcolor{LightBlue} Short Run High Gray Level Emphasis (GLRLM) & 0.180 & 0.016 \\
       Variance (First Order) & 0.161 & 0.014 \\
        \rowcolor{LightBlue} Inverse Difference Moment Normalized  (GLCM) & 0.137 & 0.016 \\
        Elongation (Shape2D) & 0.128 & 0.017 \\
        \rowcolor{LightBlue} Difference Average (GLCM) & 0.122 & 0.015 \\
      Inverse Variance (GLCM) & 0.094 & 0.016 \\
        \bottomrule
    \end{tabularx}
\end{table}

\begin{figure}[ht]
    \centering
    \includegraphics[width=1.0\linewidth]{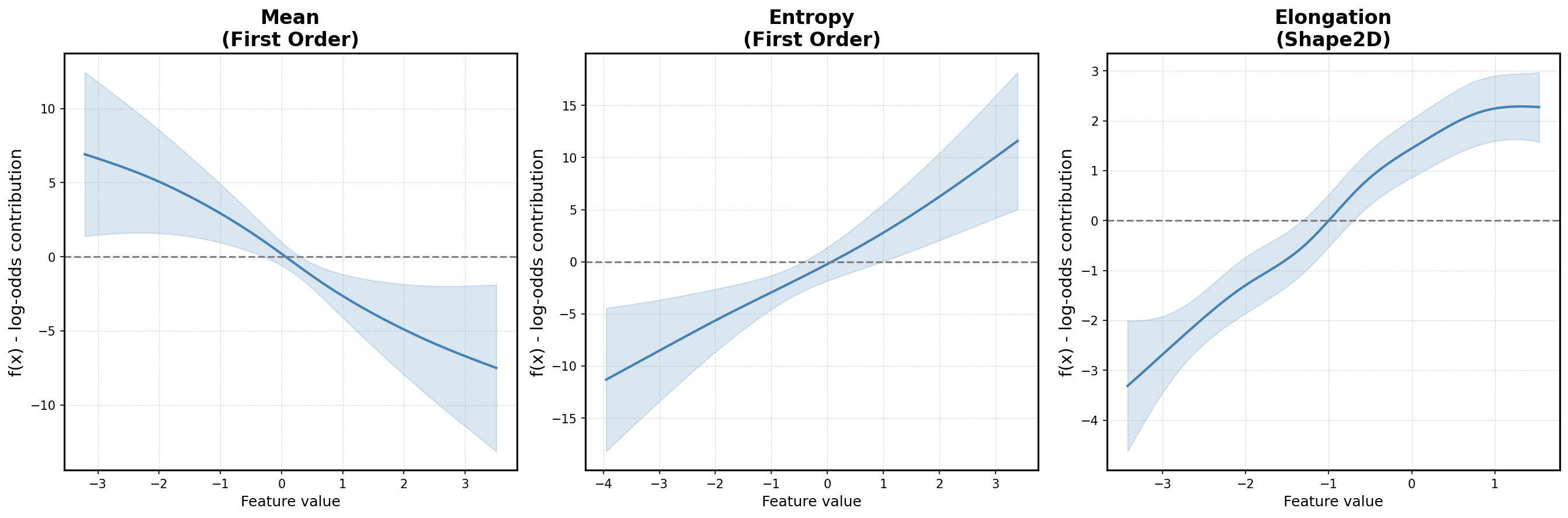}
    \caption{\textbf{Partial dependence plots of the three 
clinically interpretable features.} Each plot shows the 
GAM spline $f(x)$, the contribution of that feature 
to the log-odds of malignancy, with 95\% confidence band. 
A positive  $f(x)$ supports malignancy; a negative value supports 
benignity.}
    \label{fig:pdp}
\end{figure}

Following the clinically anchored radiomics dictionary 
of Salmanpour~\textit{et al.}~\cite{salmanpour}, three of 
our top features (\textit{Mean}, \textit{Entropy}, and 
\textit{Elongation}) could be directly linked to ACR TI-RADS 
semantic descriptors~\cite{ti_rads}, and their corresponding partial dependence plots (PDPs) 
are shown in~\autoref{fig:pdp}. \textit{Mean} maps to 
the \textit{echogenicity} descriptor: lower mean pixel intensity is associated with higher malignancy probability 
in hypoechoic nodules. \textit{Entropy} maps to the \textit{composition} descriptor: 
 solid heterogeneous nodules 
produce more complex intensity distributions, raising entropy 
and malignancy risk. \textit{Elongation} quantifies the 
\textit{taller-than-wide} shape descriptor: values above unity highlight 
vertical growth associated with higher malignancy risk.

\subsection{CCCP Results}
The results of the label-conditional cross-conformal prediction are summarized in~\autoref{tab:CCCP}. CCCP was applied with $K=10$ 
calibration folds across six significance levels in the range 
$\epsilon \in [0.05, 0.10]$.
\begin{table}[httb!]
\small
\centering
\caption{\textbf{CCCP results for GAM across $\epsilon \in [0.05, 0.10]$.}
Coverage and efficiency metrics on the test set ($n=1000$, $K=10$). Bold indicates the selected significance level.
$^\dagger$Benign coverage below target.
\label{tab:CCCP}}
    \begin{tabularx}{0.97\linewidth}{ l c c c c c c }
        \toprule
        \textbf{$\epsilon$} & \textbf{Tot. Cov.} & \textbf{Ben. Cov.} & \textbf{Mal. Cov.} & \textbf{Uncert. rate} & \textbf{Cond. NPV} & \textbf{Cond. Prec.} \\
        \midrule
        \rowcolor{LightBlue} 0.05 & 0.95 & 0.96 & 0.95 & 0.57 & 0.77 & 0.96 \\
        0.06 & 0.94 & 0.94 & 0.94 & 0.51 & 0.75 & 0.95 \\
        \rowcolor{LightBlue} 0.07 & 0.93 & 0.93 & 0.93 & 0.46 & 0.75 & 0.94 \\
        0.08 & 0.92 & 0.91 & 0.92 & 0.41 & 0.73 & 0.94 \\
        \rowcolor{LightBlue} \textbf{0.09} & \textbf{0.92} & \textbf{0.91} & \textbf{0.92} & \textbf{0.37} & \textbf{0.73} & \textbf{0.94} \\
        0.10 & 0.90 & 0.89$^\dagger$ & 0.91 & 0.33 & 0.71 & 0.93 \\
        \bottomrule
    \end{tabularx}
 
\end{table}
Among the evaluated levels, $\epsilon=0.09$ 
was the largest value in the considered range achieving label-conditional 
coverage on both classes (0.91 benign, 0.92 malignant), and was therefore 
selected as the operating significance level. At $\epsilon=0.10$, 
benign coverage dropped to 0.89, falling below target, possibly due 
to class imbalance. No empty prediction sets were observed.
As expected, the uncertainty rate decreased monotonically with 
increasing $\epsilon$, from 0.57 at $\epsilon=0.05$ to 0.33 at 
$\epsilon=0.10$. At $\epsilon=0.09$, 37\% of 
cases received a double label $\{B,M\}$ and were left to the 
clinician, while the remaining 
63\% received a certain prediction (single-label). Among certain predictions, conditional precision was consistently high ($0.93-0.96$), while conditional NPV was lower 
($0.71-0.77$), likely attributable to class imbalance. Overall, the high conditional precision indicates that the framework may reliably support biopsy recommendations for certain malignant cases, while the lower conditional NPV reflects greater difficulty in excluding malignancy.


\section{Conclusion}
\label{sec:CONCLUSIONS}
This work proposed a radiomics-based pipeline for 
thyroid nodule malignancy classification, integrating a GAM classifier, 
selected among six candidates for its best trade-off between 
discriminative performance and interpretability, with label-conditional 
cross-conformal prediction for statistically guaranteed uncertainty 
quantification. The GAM achieved competitive classification 
performance and conformal prediction provided 
clinically relevant uncertainty estimates, supporting high-confidence 
malignancy predictions while referring ambiguous cases to the clinician.

Class imbalance emerged as the main limiting factor, affecting classifier specificity and NPV, and preventing label-conditional 
coverage at $\epsilon=0.10$ for the underrepresented benign class. 
At $\epsilon=0.09$, the model achieved high conditional precision 
but lower conditional NPV, and 37\% of cases were left to the 
clinician, suggesting that the framework only partially addresses the problem of reducing unnecessary biopsies. The reliance 
on a single dataset further limited the generalizability of these results 
to different acquisition settings and patient populations. Additionally, 
radiomic features were extracted from bounding box crops rather than precise nodule segmentations, which may introduce noise from surrounding 
thyroid tissue and limit feature accuracy.

As a work in progress, future developments will focus on addressing class imbalance both at the classifier training stage and through conformal prediction methods more robust to skewed classes. Further directions will include multi-dataset validation, integration of nodule segmentation prior to feature extraction, and deeper clinical interpretation of radiomic features.

\section*{Conflict of interests}
\label{sec:CONFLICT-OF-INTERESTS}
The authors have no conflicts of interest to disclose.


\section*{Availability of data and software code}
\label{sec:AVAILABILITY}
Our software code is available at the following URL:

\url{https://github.com/laura-testa/thyroid-radiomics-cp}
\newline
\indent The TN5000 dataset is available at the following URL: 

\url{https://figshare.com/s/cb6a67f17c04b29e7edd}.

\footnotesize
\bibliographystyle{unsrt}
\bibliography{bibliography_CIBB_file.bib} 
\normalsize

\end{document}